\documentclass[prl,showpacs,amsmath,amssymb]{revtex4}
\usepackage{graphicx}

\begin{document}

\title{Collective Quantization of a Gravitating Skyrmion}

\author{Noriko Shiiki}
\email{norikoshiiki@mail.goo.ne.jp}
\author{Nobuyuki Sawado}
\email{sawado@ph.noda.tus.ac.jp}
\author{Shinsho Oryu}
\email{oryu@ph.noda.tus.ac.jp}

\affiliation{Department of Physics, Tokyo University of Science, Noda, Chiba 278-8510, Japan}

\date{\today}

\begin{abstract}
Collective quantization of a $B=1$ gravitating skyrmion is described. 
The rotational and isorotational modes are quantized in the same 
manner as the skyrmion without gravity. 
It is shown in this paper how the static properties of nucleons such as masses, 
charge densities, magnetic moments are modified by the gravitational 
interaction.   
\end{abstract}

\pacs{12.39.Dc, 21.10.-k, 04.20.-q}
\maketitle 

\section{1. Introduction}
The Skyrme model is a nonlinear meson theory proposed by T. H. R. Skyrme~\cite{skyrme58}. 
It gives a unified description of hadronic physics by incorporating baryons as 
topological solitons. The baryon number $B$ corresponds to the topological charge. 
Following the findings of the Finkelstein-Rubinstein constraints  
which enable a single skyrmion to be quantized as a fermion~\cite{finkelstein68}, 
the model was identified with QCD in the Large-$N_{c}$ limit by E. Witten~\cite{witten79}.  
The static properties of nucleons such as masses, mean radius, charge densities 
and magnetic moments are evaluated upon collective (zero mode) quantization 
of the skyrmion~\cite{adkins83, brown86}. 

The Einstein-Skyrme system has been studied by several authors. 
The first obtained solutions in this system are spherically symmetric black holes 
with Skyrme hair~\cite{luckock86,luckock87,droz91}. 
It was the first counter example of the no-hair conjecture for black holes discovered. 
Later, regular solutions for $B=1$~\cite{luckock87,droz91,bizon92} and axially 
symmetric black hole and regular solutions for $B=2$~\cite{shiiki02} were found. 
The extended models to $SU(3)$ and $SU(N)$ were also studied in Refs.~\cite{zakrzewski04}.     

It is, however, necessary to quantize those skyrmions to interpret as gravitating or 
black hole nucleons.  Therefore, in this paper we shall perform collective quantization 
of the $B=1$ gravitating skyrmion and compute its static properties 
following the work of Ref.~\cite{adkins83}. 
The effects of gravity on the nucleon observables are examined.  
Since the skyrmion picture of a nucleon is correct within about $30\%$ error, 
it is difficult to estimate the effects quantitatively. 
However, we believe that the results we have obtained can help us to 
understand qualitatively the effects of gravity on the nucleon properties. 

In the Einstein-Skyrme theory, the Planck mass is related to the pion decay 
constant $F_{\pi}$ and the coupling constant $\alpha$ by $M_{pl}=F_{\pi}\sqrt{4\pi/\alpha}$. 
To realize the realistic value of the Planck mass, the coupling constant should be extremely 
small with $\alpha \sim O(10^{-39})$, which makes the theory little different from the theory 
without gravity. 
However, some theories such as scalar-tensor gravity theory~\cite{brans61} 
and theories with extra dimensions predict the time variation of the gravitational 
constant~\cite{marciano84}. Thus there may have been an epoch in the early universe 
where the gravitational effects on nucleons were significant. We consider those 
effects worth being studied in the Skyrme model. 

It is discussed that skyrmions could be produced in a manner analogous to 
the production of cosmic strings and monopoles in the early universe via the Kibble 
mechanism~\cite{turok89}. This mechanism has been applied to the study of the 
production of baryons/antibaryons in jet events~\cite{ellis88} and in quark gluon 
plasma~\cite{degrand84,ellis89}. 
It may be possible to extend our work to these interesting high-energy phenomena 
where the hot and dense conditions in the early universe are mimicked. 

\section{2. Classical Gravitating Skyrmions}
The classical regular solutions of the Einstein-Skyrme system with $B=1$ have been 
already studied in Refs.~\cite{luckock87,droz91,bizon92}. 
We therefore give a short review of the model and solutions in this section. 

The Skyrme model coupled with gravity can be defined by the Lagrangian
\begin{eqnarray}
	{\cal L}={\cal L}_{G}+{\cal L}_{S} \label{lagrangian}
\end{eqnarray}
where 
\begin{eqnarray}
	{\cal L}_{G}&=&\frac{1}{16\pi G}R \label{lg} \\
	{\cal L}_{S}&=&-\frac{F_{\pi}^2}{16}g^{\mu\nu}{\rm tr}(\partial_{\mu}U\partial_{\nu}U^{-1})
	+\frac{1}{32e^{2}}g^{\mu\nu}g^{\rho\sigma}{\rm tr}[(\partial_{\mu}U)U^{-1},(\partial_{\rho}U)U^{-1}]
	[(\partial_{\nu}U)U^{-1},(\partial_{\sigma}U)U^{-1}] \label{ls}
\end{eqnarray}
where $U$ is an $SU(2)$-valued chiral field, $F_{\pi}$ is the pion decay constant and 
$e$ is a dimensionless free parameter. 
The $B=1$ skyrmion can be obtained by imposing the hedgehog ansatz on the chiral field  
\begin{eqnarray}
	U=\cos F(r) + i{\vec n}\cdot {\vec \tau}\sin F(r) \, . \label{hedgehog}
\end{eqnarray}
Correspondingly we consider the static spherically symmetric metric given by  
\begin{eqnarray}
	ds^{2}=-N^{2}(r)C(r)dt^{2}+\frac{1}{C(r)}dr^{2}+r^{2}d\Omega^{2} \label{metric}
\end{eqnarray}
where we have defined 
\begin{eqnarray*}
	C(r)=1-\frac{2m(r)}{r}\, .
\end{eqnarray*}
Inserting these ansatz into the Lagrangian (\ref{ls}), one obtains the static 
energy density for the chiral field 
\begin{eqnarray}
	{\cal E}_{S}=\frac{F_{\pi}^{2}}{8}\left(CF'^{2}+\frac{2\sin^{2}F}{r^{2}}\right)
	+\frac{1}{2e^{2}}\frac{\sin^{2}F}{r^{2}}\left(2CF'^{2}+\frac{\sin^{2}F}{r^{2}}\right) \, . 
	\label{}
\end{eqnarray}
Let us introduce dimensionless variables   
\begin{eqnarray*}
	x=eF_{\pi}r \; , \;\;\; \mu (x)=eF_{\pi}m(r)\, . 
\end{eqnarray*}
In terms of $x$ and $\mu$, the static energy thus can be written by 
\begin{eqnarray}
	E_{S}=4\pi \frac{F_{\pi}}{e} \int \left\{\frac{1}{8}\left(CF'^{2}
	+\frac{2\sin^{2}F}{x^{2}}\right)+\frac{\sin^{2}F}{2x^{2}}
	\left(2CF'^{2}+\frac{\sin^{2}F}{x^{2}}\right)\right\}N x^{2} dx \, . \label{energy}
\end{eqnarray}
The covariant topological current is defined by  
\begin{eqnarray}
	B^{\mu}=-\frac{\epsilon^{\mu\nu\rho\sigma}}{24\pi^{2}}\frac{1}{\sqrt{-g}}
	{\rm tr}\left(U^{-1}\partial_{\nu}UU^{-1}\partial_{\rho}UU^{-1}\partial_{\sigma}U\right). 
	\label{topological_current} \label{baryon_current}
\end{eqnarray}
whose zeroth component corresponds to the baryon number density 
\begin{eqnarray}
	B^{0}=-\frac{1}{2\pi^{2}}\frac{1}{N}\frac{F'\sin^{2}F}{r^{2}}\, . \label{}
\end{eqnarray}
Topological soliton solutions can be obtained if the following boundary conditions 
for the profile function are considered
\begin{eqnarray}
	F(0)=k\pi \;, \;\;\; F(\infty)=0   \label{}
\end{eqnarray}
where $k$ is an arbitrary integer. Then the baryon number becomes    
\begin{eqnarray}
	B=\int \sqrt{-g}\,B^{0} \, d^{3}x = -\frac{2}{\pi}\int_{k\pi}^{0}
	\sin^{2}F dF = k \, . \label{}
\end{eqnarray}
Since our concern is a $B=1$ skyrmion, $k$ is restricted to be one hereafter.   

The field equations for the gravitational fields $N(x)$ and $\mu (x)$ can be 
derived from the Einstein equations as 
\begin{eqnarray}
	N'&=& \frac{\alpha}{4}\left(x+\frac{8\sin^{2}F}{x}\right)NF'^{2} \\
	\mu'&=& \frac{\alpha}{8} \left[(x^{2}+8\sin^{2}F)CF'^{2}
	+2\sin^{2}F+\frac{4\sin^{4}F}{x^{2}}\right]
\end{eqnarray} 
where we have defined the coupling constant $\alpha = 4\pi GF_{\pi}^{2}$. 
The variation of the static energy (\ref{energy}) with respect to the profile $F(x)$ 
leads to the field equation for matter 
\begin{eqnarray}
	F''&=& \frac{1}{NC(x^{2}+8\sin^{2}F)}\left[-(x^{2}+8\sin^{2}F)N'CF'
	+\left(1+\frac{4\sin^{2}F}{x^{2}}+4CF'^{2}\right)N\sin 2F \right. \nonumber \\
	&& \left. -2(x+4\sin 2FF')NCF'-2\left(1+\frac{8\sin^{2}F}{x^{2}}\right)
	(\mu-\mu'x)NF' \right] \, .\label{}
\end{eqnarray}
To solve these coupled field equations, let us consider the boundary conditions for the 
gravitational fields. Expanding the fields $F(x),\mu(x),N(x)$ around the origin and 
substituting into the field equations, one obtains 
\begin{eqnarray*}
	F(x)&=&\pi+b_{1}x+O(x^{3}) \\
	\mu(x)&=&\frac{\alpha}{8} b_{1}^{2}(1+4b_{1}^{2})x^{3}+O(x^{4}) \\
      N(x)&=& b_{2}+\frac{\alpha}{4}b_{1}^{2}b_{2}
	(1+8b_{1}^{2})x^{2}+O(x^{3}) \end{eqnarray*}
where $b_{1}$ and $b_{2}$ are shooting parameters determined so as to satisfy the boundary 
conditions at infinity $F(\infty)=0$ and $N(\infty)=1$. 

\begin{figure}
\includegraphics[height=6.5cm, width=9cm]{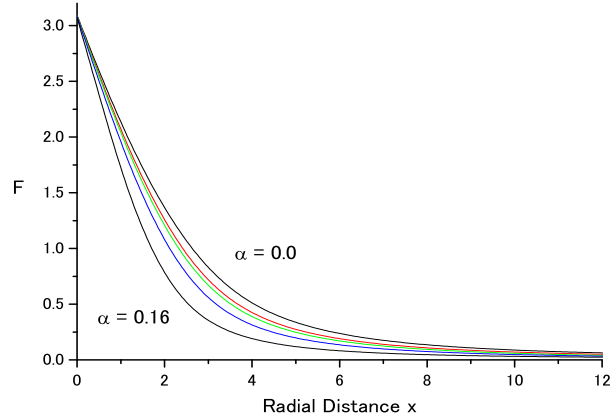}
\caption{\label{fig:profile} Radial dependence of the profile function with 
$\alpha = 0.0, 0.04, 0.08, 0.12, 0.16$ respectively. }
\end{figure}

The skyrmion solutions for the various values of the coupling constant are shown in 
Fig.~\ref{fig:profile}. As is shown in Ref.~\cite{bizon92}, there exist two branches 
of solutions depending on the stability. 
We have examined only the solution in the stable branch since it is physically interesting 
as a nucleon.  In particular, the stable solution with $\alpha = 0.0$ recovers the solution 
in flat spacetime and accord with the solution obtained in Ref.~\cite{adkins83}. 
No solution exists for $\alpha \gtrsim 0.162$. 

\section{3. Collective Quantization}

\begin{figure}
\includegraphics[height=6.5cm, width=9cm]{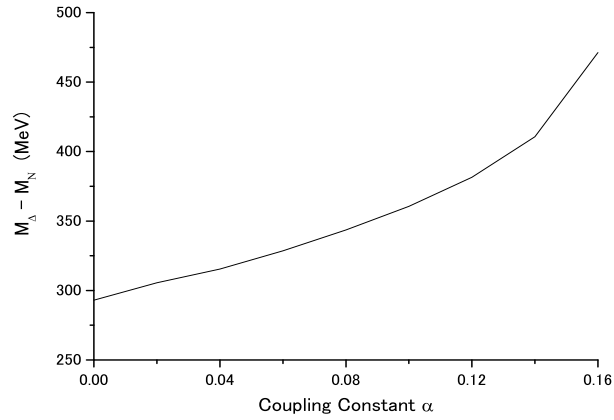}
\caption{\label{fig:mass_diff} Coupling constant dependence of the $N-\Delta$ 
mass difference in units of MeV. }
\end{figure}

To describe physical nucleon and $\Delta$ states, we need to perform 
quantization for the classical skyrmion. 
The field theory is truncated to certain collective degrees of freedom 
of the skyrmion, which reduces the problem to a simple quantum 
mechanics on the collective space. 
From the symmetry of the Lagrangian, one can see that if $U$ is the soliton 
solution, then $U\rightarrow AUA^{-1}$ where $A$ is an arbitrary constant $SU(2)$ matrix, 
is also a solution with the same finite energy. Therefore $A$ is the collective 
coordinate to be quantized. Letting the matrix $A$ time dependent $A(t)$, 
we replace the field $U$ in the Lagrangian (\ref{ls}) as 
\begin{eqnarray}
	U({\vec r},t)=A(t)U_{0}A(t)^{-1} \label{ut}
\end{eqnarray}
where $U_{0}$ is the hedgehog solution constructed in the previous section.     
For the hedgehog solution, the spin and isospin rotation are equivalent. 
Definite spin and isospin states are obtained by quantizing those degrees of freedom. 
Substituting the transformation of the chiral fields (\ref{ut}) into (\ref{ls}), 
one can get  
\begin{eqnarray}
	L_{S}=-M_{B=1}+\lambda \, {\rm tr}({\dot A}{\dot A}^{-1})
	=-M_{B=1}+2\lambda \sum_{i=0}^{3}{\dot a}^{2}  \label{}
\end{eqnarray}
where
\begin{eqnarray}
	\lambda=\frac{2\pi}{3F_{\pi}e^{3}}\Lambda \; , \;\;\;
	\Lambda=\int_{0}^{\infty} \frac{1}{NC}\left[1+4\left(CF'^{2}+\frac{\sin^{2}F}{x^{2}}
	\right)\right]x^{2}\sin^{2}F\, dx
\end{eqnarray} 
and $M_{B=1}$ is the $B=1$ classical skyrmion mass and in the second equality, 
we have parameterized $A=a_{0}+i{\vec \tau}\cdot {\vec a}$ with $a_{0}^{2}+{\vec a}^{2}=1$. 
Canonical quantization can be performed in standard manners in terms of $a$'s.  
The Hamiltonian is then diagonalized as 
\begin{eqnarray}
	H=M_{B=1}+\frac{1}{8\lambda}\sum_{i=0}^{3}\left(-\frac{\partial^{2}}{\partial a_{i}^{2}}
	\right)=M_{B=1}+\frac{l(l+2)}{8\lambda} \label{hamiltonian}
\end{eqnarray}
where $l=2I=2J$, and $(I,J)$ are respectively the isospin and spin quantum number 
with the operators 
\begin{eqnarray}
	I_{k}=\frac{i}{2}\left(a_{0}\frac{\partial}{\partial a_{k}}
	-a_{k}\frac{\partial}{\partial a_{0}}-\epsilon_{klm}a_{l}
	\frac{\partial}{\partial a_{m}}\right)\; , \;\;\; 
	J_{k}=\frac{i}{2}\left(a_{k}\frac{\partial}{\partial a_{0}}-a_{0}
	\frac{\partial}{\partial a_{k}}-\epsilon_{klm}a_{l}
	\frac{\partial}{\partial a_{m}}\right)\, . \label{}
\end{eqnarray}
Let us note that if the scaling parameter is explicitly written in the Hamiltonian 
(\ref{hamiltonian}), one can see that this quantization corresponds to the expansion 
around the skyrmion in powers of $1/N_{c}$ with the second term being of order $1/N_{c}$. 
The nucleon, delta mass and their mass difference are thus given by 
\begin{eqnarray}
	M_{N}=M_{B=1}+\frac{1}{2\lambda}\frac{3}{4} \; , \;\;\;
	M_{\Delta}=M_{B=1}+\frac{1}{2\lambda}\frac{15}{4}\; , \;\;\;
	M_{\Delta}-M_{N}=\frac{3}{2\lambda}\, . \label{}
\end{eqnarray}
For higher $l$ states, there are no counterparts in nature and they are considered 
as artifacts of this model. Hereafter we adopt the same parameter set as 
in Ref.~\cite{adkins83} $F_{\pi}=129$MeV and $e=5.45$ so that for $\alpha=0.0$, 
the experimental values of a nucleon and delta mass are reproduced with about $30\%$ error. 
Fig.~\ref{fig:mass_diff} shows the $\alpha $ dependence of the mass difference 
between $N$ and $\Delta$ in units of MeV. It is seen that the mass difference 
increases monotonically with increasing $\alpha$. Fig.~\ref{fig:profile} 
implies that the strong gravity makes the size of the skyrmion smaller which 
makes the inertial moment smaller, resulting in increase in the mass difference. 
In the collective quantization, the skyrmion can be quantized as a slowly rotating 
rigid body and the mass difference between the delta and nucleon is interpreted 
as a consequence of the rotational kinetic energy. 
Thus the gravity works for increasing the kinetic energy of the skyrmion. 
In the naive $SU(6)$ quark model, the mass difference is ascribed to 
the hyperfine splittings. The increase in the mass difference  
may imply that due to the reduction of the distance between quarks, 
the effects of the hyperfine splittings become dominant by the gravity~\cite{glashow75}.   

\begin{figure}
\includegraphics[height=6.5cm, width=9cm]{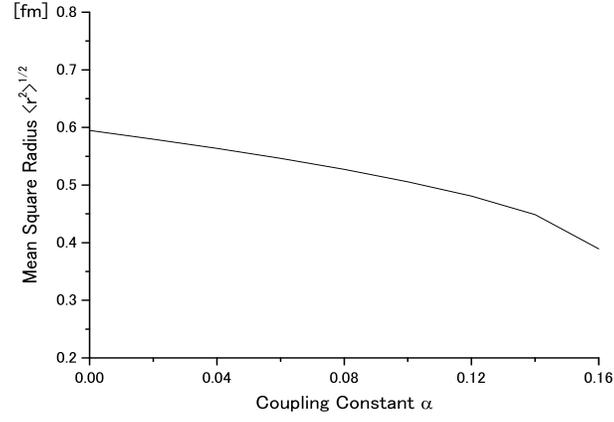}
\caption{\label{fig:mean_radius} Coupling constant dependence of the mean square radius 
in units of fm. }
\end{figure}

The isoscalar mean square radius of the nucleon is defined in terms of the baryon number 
density by   
\begin{eqnarray}
	\langle r^{2}\rangle =\int \sqrt{-g}\,r^{2}B^{0}(r)\, d^{3}x
	 = -\frac{1}{(eF_{\pi})^{2}}\frac{2}{\pi}\int_{0}^{\infty} 
	x^{2}F'\sin^{2}F\,dx \, . \label{}
\end{eqnarray}
Fig.~\ref{fig:mean_radius} shows the $\alpha$ dependence of the root mean square radius. 
It decreases with increasing $\alpha$, which confirms the attractive effect of the gravity. 

\begin{figure}
\includegraphics[height=6.5cm, width=9cm]{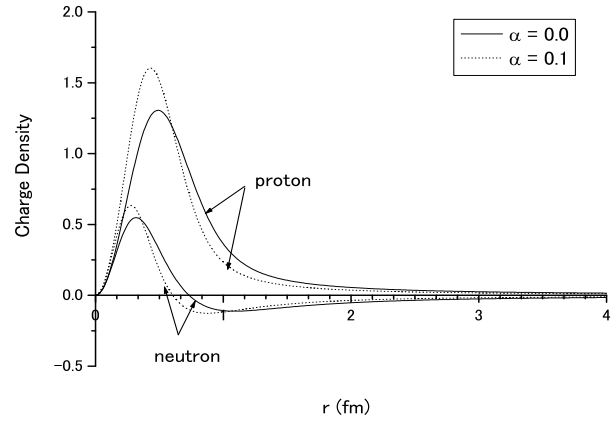}
\caption{\label{fig:charge_density} Radial dependence of proton and neutron 
charge densities as functions of the radial distance $r$ (fm). }
\end{figure}

To compute the charge densities and magnetic moments, let us derive the baryon and isovector current. 
From Eq.~(\ref{topological_current}), the baryon current can be written by   
\begin{eqnarray}
	B^{i}=-\frac{\epsilon^{ijk}}{8\pi^{2}}\frac{1}{\sqrt{-g}}{\rm tr}
	(L_{j}L_{k}L_{0})   \label{baryon_current}
\end{eqnarray}
where $L_{i}=U^{-1}\partial_{i}U$. 
Substituting (\ref{hedgehog}) and (\ref{ut}) into (\ref{baryon_current}), 
one can get 
\begin{eqnarray}
	B^{i}=i\frac{\epsilon^{ijk}}{2\pi^{2}}\frac{1}{\sqrt{-g}}\frac{\sin^{2}F}{r}
	F'{\hat r}^{j}\,{\rm tr}\,(\tau^{k}{\dot A}^{-1}A) \label{}
\end{eqnarray}
where ${\hat r}^{j}=x^{j}/r$ and we have used the identity 
\begin{eqnarray}
	{\dot A}^{-1}A=\frac{\tau^{a}}{2}{\rm tr}(\tau^{a}{\dot A}^{-1}A)\, . \label{}
\end{eqnarray}
From the Skyrme Lagrangian (\ref{ls}), one can construct the Noether current 
for the $SU(2)_{L}$ transformation $\delta U=iQ_{L}U$ as 
\begin{eqnarray}
	J_{L}^{\mu}=-\frac{iF_{\pi}^{2}}{8}g^{\mu\nu}{\rm tr}(Q_{L}R_{\nu})
	+\frac{i}{8e^{2}}g^{\mu\nu}g^{\rho\sigma}{\rm tr}([Q_{L},R_{\rho}][R_{\nu},R_{\sigma}])  \label{}
\end{eqnarray}
where $R_{\mu}=U\partial_{\mu}U^{-1}=-(\partial_{\mu}U)U^{-1}$ is the right current. 
Similarly, for $SU(2)_{R}$ transformation $\delta U=iUQ_{R}$, one obtains  
\begin{eqnarray}
	J_{R}^{\mu}=\frac{iF_{\pi}^{2}}{8}g^{\mu\nu}{\rm tr}(Q_{R}L_{\nu})-\frac{i}{8e^{2}}
	g^{\mu\nu}g^{\rho\sigma}{\rm tr}[Q_{R},L_{\rho}][L_{\nu},L_{\sigma}] \label{}
\end{eqnarray}
where $L_{\mu}=U^{-1}\partial_{\mu}U$ is the left current. 
The relations between vector and axial transformations and left and right transformations  
\begin{eqnarray}
	Q_{V}=\frac{1}{2}(Q_{L}+Q_{R}) \; ,\;\;\; Q_{A}=\frac{1}{2}(Q_{R}-Q_{L}) \label{}
\end{eqnarray}
lead to the vector and axial currents
\begin{eqnarray}
	V^{\mu,a}&=&\frac{iF_{\pi}^{2}}{8}g^{\mu\nu}\,{\rm tr}\,(Q^{a}(R_{\nu}+L_{\nu}))
	+\frac{i}{8e^{2}}g^{\mu\nu}g^{\rho\sigma}\,{\rm tr}\,(Q^{a}[R_{\rho},[R_{\nu},R_{\sigma}]]
	+Q^{a}[L_{\rho},[L_{\nu},L_{\sigma}]]) \\
	A^{\mu,a}&=&\frac{iF_{\pi}^{2}}{8}g^{\mu\nu}\,{\rm tr}\,(Q^{a}(R_{\nu}-L_{\nu}))
	+\frac{i}{8e^{2}}g^{\mu\nu}g^{\rho\sigma}\,{\rm tr}\,(Q^{a}[R_{\rho},[R_{\nu},R_{\sigma}]]
	-Q^{a}[L_{\rho},[L_{\nu},L_{\sigma}]])\, . \label{}
\end{eqnarray}
For convenience, we perform the integration of the vector and axial current to obtain 
\begin{eqnarray}
      && \int \sqrt{-g}\,V^{0,a} \,d^{3}x=i\frac{F_{\pi}}{e}\frac{2\pi}{3}\Lambda \,{\rm tr}
      \,(\tau^{a}{\dot A}A^{-1})\\
	&& \int \sqrt{-g}\,{\vec q}\cdot {\vec r}\, V^{i,a} \,d^{3}x= 
	-\frac{F_{\pi}}{e}\frac{\pi}{3}\Sigma \,q^{l}\epsilon_{lim}{\rm tr}\,(\tau^{m}A^{-1}
	\tau^{a}A) \\ 
	&& \int \sqrt{-g}\,A^{i,a} \,d^{3}x = \frac{1}{e^{2}}\frac{\pi}{3}D \,{\rm tr}
	\,(\tau^{a}A\tau^{i}A^{-1}) \label{}
\end{eqnarray}
where 
\begin{eqnarray}
      \Sigma &=& \int Nx^{2}\sin^{2}F\left[1+4\left(CF'^{2}+\frac{\sin^{2}F}{x^{2}}\right)\right] dx \\
	D&=& \int Nx^{2}\left[CF'+\frac{\sin 2F}{x}+4\left(C\frac{\sin 2F}{x}F'^{2}
	+2C\frac{\sin^{2}F}{x^{2}}F'+\frac{\sin^{2}F\sin 2F}{x^{3}}\right)\right] dx \, .\label{}
\end{eqnarray}
The isoscalar and isovector charge densities per unit $r$ are then given by 
\begin{eqnarray}
	\rho_{I=0}&=&\int \sqrt{-g}\,B^{0} d\theta d\varphi =-eF_{\pi}\frac{2}{\pi}\sin^{2}FF' \\
	\rho_{I=1}&=&eF_{\pi}Nx^{2}\sin^{2}F\left[1+4
      \left(CF'^{2}+\frac{\sin^{2}F}{r^{2}}\right)\right]/\Sigma \, .\label{}
\end{eqnarray}
The charge density for the proton and neutron are given by 
\begin{eqnarray}
	\rho_{p}=\frac{1}{2}(\rho_{I=0}+\rho_{I=1}) \; ,\;\;\;
	\rho_{n}=\frac{1}{2}(\rho_{I=0}-\rho_{I=1})  \label{}
\end{eqnarray}
and are shown in Fig.~\ref{fig:charge_density} for $\alpha = 0.0,\, 0.1$. 
For the strong gravity, the peaks of the charge densities become higher 
and move towards the center. 

\begin{figure}
\includegraphics[height=6.5cm, width=9cm]{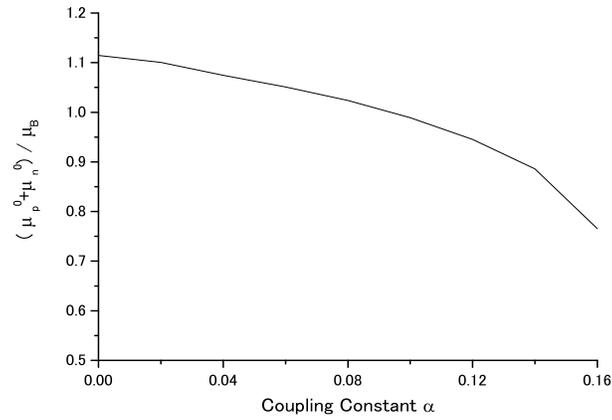}
\caption{\label{fig:isoscalar_mag} Coupling constant dependence of the isoscalar 
magnetic moments $\mu_{p}^{0}+\mu_{n}^{0}$ in units of Bohr magnetons $\mu_{B}$. }
\end{figure}

\begin{figure}
\includegraphics[height=6.5cm, width=9cm]{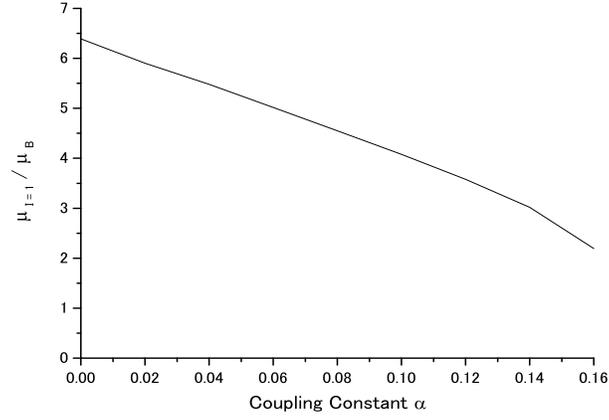}
\caption{\label{fig:vector_mag} Coupling constant dependence of the isovector 
magnetic moment in units of Bohr magnetons $\mu_{B}$. }
\end{figure}

The isoscalar and isovector magnetic moments in the nucleon rest frame are expressed 
in terms of the baryon and isovector current as  
\begin{eqnarray}
	{\vec \mu}_{I=0}&=&\frac{1}{2}\int \sqrt{-g}\; {\vec r}\times {\vec B}\,d^{3}x \\ 
	{\vec \mu}_{I=1}&=&\frac{1}{2}\int \sqrt{-g}\; {\vec r}\times {\vec V}^{3}\,d^{3}x \, . \label{}
\end{eqnarray}
The expectation value of the isoscalar magnetic moments in a proton spin-up 
state are thus given by  
\begin{eqnarray}
	\mu_{p}^{0}=\langle p\uparrow|(\mu_{I=0})_{i}|p\uparrow \rangle=\frac{i}{4\pi^{2}}
	\int({\hat r}^{i}{\hat r}^{j}-\delta^{ij})\sin^{2}FF'\, d^{3}x 
	\,{\rm tr}\,(\tau^{j}{\dot A}^{-1}A)=\frac{1}{4\pi}\frac{e}{F_{\pi}}
	\frac{\langle x^{2}\rangle}{\Lambda}\delta^{i3}\, . \label{}
\end{eqnarray}
The neutron isoscalar magnetic moment is equal to the proton isoscalar magnetic moment.  
Fig.~\ref{fig:isoscalar_mag} shows the third component of the isoscalar magnetic moment 
measured in Bohr magnetons $\mu_{B}=1/2M_{N}$. It decreases with increasing $\alpha$. 
The isovector magnetic moment can be obtained in a similar manner as of the isoscalar 
\begin{eqnarray}
	\mu_{p}^{1}&=& \langle {\rm p}\uparrow | \int \sqrt{-g} \, \epsilon_{li3}\,x^{l} V^{i,3}\, 
	 d^{3}x \,|{\rm p}\uparrow \rangle \\ 
	&=& -\frac{\pi}{3}\frac{\Sigma}{e^{3}F_{\pi}} \langle {\rm p}\uparrow |
	{\rm tr}\,(\tau^{3}A^{-1}\tau^{3}A)|{\rm p}\uparrow\rangle  \\
	&=& \frac{2\pi}{9}\frac{\Sigma}{e^{3}F_{\pi}}\label{}
\end{eqnarray}
where we have used the relation for the third equality 
\begin{eqnarray}
	 \langle {\rm p}\uparrow |
	{\rm tr}\,(\tau^{i}A^{-1}\tau^{j}A)|{\rm p}\uparrow\rangle =-\frac{2}{3}
	\langle {\rm p}\uparrow |\sigma^{i}\tau^{i}|{\rm p}\uparrow\rangle \,. \label{}
\end{eqnarray}
The neutron isovector magnetic moment has the same value but opposite sign. 
Thus the isovector magnetic moment is defined by 
$\mu_{I=1}=\mu_{p}^{1}-\mu_{n}^{1}=2\mu_{p}^{1}$.  
Fig.~\ref{fig:vector_mag} shows the computed isovector magnetic moment. 
As in the isoscalar moment, it decreases with increasing $\alpha$ but 
the effect of gravity is more evident.  

\begin{figure}
\includegraphics[height=6.5cm, width=9cm]{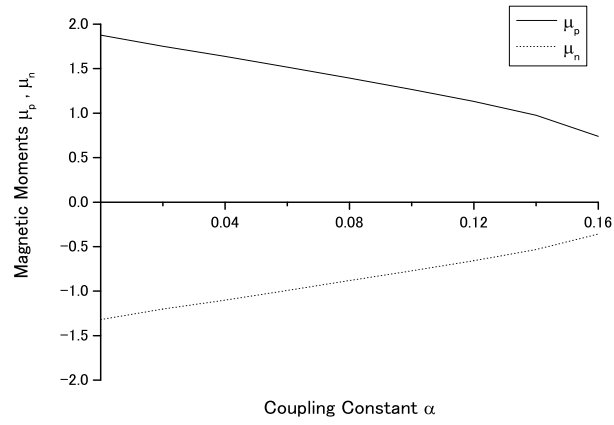}
\caption{\label{fig:total_mag} Coupling constant dependence of the proton and 
neutron magnetic moment in units of Bohr magnetons $\mu_{B}$. }
\end{figure}

The magnetic moments for the proton and neutron are given by the sum of the 
isoscalar and vector magnetic moment 
\begin{eqnarray}
	\mu_{p\,(n)}=\frac{1}{2\mu_{B}}(\mu_{p\,(n)}^{0}+\mu_{p\,(n)}^{1})\, . \label{}
\end{eqnarray}
Fig.~\ref{fig:total_mag} shows the $\alpha$ dependence of the proton and 
neutron magnetic moments. 
The absolute values are both decreasing with increasing $\alpha$. 
This result indicates that although in the situation where gravitational 
effects are negligible, the assumption that the observed baryons are 
three-quark states with zero orbital angular momentum is a good approximation, 
it may no longer be valid under the strong gravitational field. 
In this case, the ground states of strongly interacting systems are not in 
pure S-wave and other states with non-zero orbital angular momenta should 
be taken into account.      

\begin{figure}
\includegraphics[height=6.5cm, width=9cm]{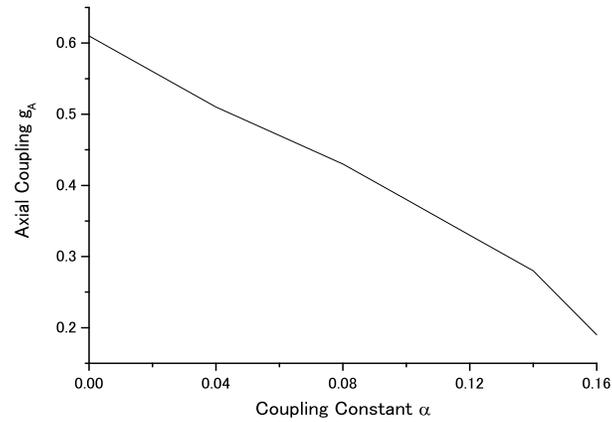}
\caption{\label{fig:axial_g} Coupling constant dependence of the axial coupling $g_{A}$. }
\end{figure}

The axial coupling can be computed from the integral of the axial current 
\begin{eqnarray}
	g_{A}=-\frac{\pi}{3e^{2}}D \, . \label{}
\end{eqnarray}
The result is shown in Fig.~\ref{fig:axial_g}. 
Since the axial coupling is related to $g_{\pi NN}$ and $g_{\pi\Delta N}$, 
our result implies that the strong gravity reduces the baryon decay rate, 
stabilizing the baryon against the strong interaction. 

\begin{figure}
\includegraphics[height=6.5cm, width=9cm]{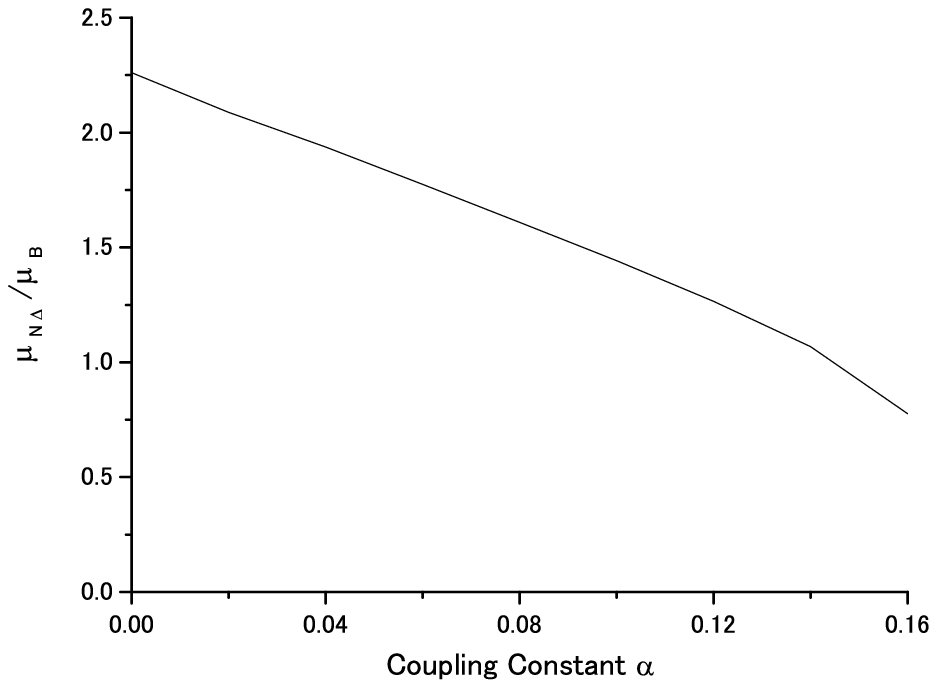}
\caption{\label{fig:transition_moment} Coupling constant dependence of the transition 
moment between $\Delta$ and $N$ via electromagnetic interaction. }
\end{figure}

In Ref.~\cite{beg64}, it was shown using a simple quark model that all 
allowed transition moments between octet and decuplet can be expressed 
in terms of the proton magnetic moment.  
Based on this argument, Adkins {\it et al.} derived the relation between the transition 
moment of $\Delta \rightarrow N\gamma$ and the proton, neutron magnetic moments 
in Ref.~\cite{adkins83} which is given by 
\begin{eqnarray}
	\mu_{N\Delta}=\sqrt{\frac{1}{2}}(\mu_{p}-\mu_{n}) \,. \label{}
\end{eqnarray}
Assuming that this derived relation is also valid under the influence of gravity, 
we have computed the transition moment and shown in Fig.~\ref{fig:transition_moment}. 
It decreases with increasing of $\alpha$ as expected from the results of the magnetic moments.   
Since the strong gravity reduces the transition moment significantly, it may be possible 
to determine the gravitational constant by observing the variation in $\mu_{N\Delta}$. 
It is interesting that the decay rates are reduced by the gravitational 
effects whether the interaction is strong or electromagnetic, which means 
the gravity works as a stabilizer of baryons. 
  
\section{4. Conclusions} 

We have performed collective quantization of a $B=1$ skyrmion in the 
Einstein-Skyrme system and investigated the static nucleon properties. 
Although the spacetime is curved, the collective space remains $SU(2)$. 
Therefore the quantization is rather straightforward. 
Modification by the gravitational interaction appears in the observables 
such as N-$\Delta$ mass difference, mean radius, charge densities, 
magnetic moments, transition moments. 
The qualitative change in the mass difference, mean square radius 
and charge densities under the strong gravitational influence 
confirm the attractive feature of the gravity. 
On the other hand, the reduction of the axial coupling and transition moments 
by the strong gravity indicate the gravitational effects as a stabilizer of baryons. 
Although the Skyrme model describes a nucleon with about $30\%$ error, 
the possibility that it may provide qualitatively correct description 
of the interaction of a nucleon with gravity can not be excluded. 
It is expected that in the early universe or equivalent high energy 
experiments, the gravitational interaction with nucleons is not negligible. 
We hope that our work could provide insight into the observations in such situations. 
It will be also interesting to quantize gravitating skyrmions with higher 
baryon numbers or black hole skyrmions in future.

\section{Acknowledgement}
We would like to thank Rajat K. Bhaduri for drawing our attention to this subject and 
useful comments.

\begin{thebibliography}{99}

\bibitem{skyrme58}
T. H. R. Skyrme, Proc. Roy. Soc. A260 (1961) 127. 

\bibitem{finkelstein68}
D. Finkelstein and J. Rubinstein, J. Math. Phys. 9 (1968) 1762.

\bibitem{witten79}
E. Witten, Nucl. Phys. B160 (1979) 57; Nucl. Phys. B223 (1983) 422; 
Nucl. Phys. B223 (1983) 433.

\bibitem{adkins83}
G. S. Adkins, C. R. Nappi and E. Witten, Nucl. Phys. B228 (1983) 552. 

\bibitem{brown86}
I. Zahed and G. E. Brown, Phys. Rept. 142 (1986) 1. 

\bibitem{luckock86}
H. Luckock and I. G. Moss, Phys. Lett. B176 (1986) 341. 

\bibitem{luckock87}
H. Luckock, "String Theory, Quantum Cosmology and Quantum Gravity, 
Integrable and Conformal Invariant Theories", Edited by de Vega 
and N. Sanchez (World Scientific, Singapore, 1987). 

\bibitem{droz91}
S. Droz, M. Heusler and N. Straumann, Phys. Lett. B268 (1991) 371. 

\bibitem{bizon92}
P. Bizon and T. Chmaj, Phys. Lett. B297 (1992) 55. 

\bibitem{shiiki02}
N. Sawado and N. Shiiki, gr-qc/0307115; Gen. Rel. Grav. 36 (2004) 1361. 

\bibitem{zakrzewski04}
B. Kleihaus, J. Kunz and A. Sood, Phys. Lett. B352 (1995) 247; 
Y. Brihaye, B. Hartmann, T. Ioannidou and W. Zakrzewski, Phys. Rev. D69 (2004) 124035; 
T. Ioannidou, B. Kleihaus and W. Zakrzewski, gr-qc/0407035. 

\bibitem{brans61}
C. Brans and R. H. Dicke, Phys. Rev. 124 (1961) 925. 

\bibitem{marciano84}
W. J. Marciano, Phys. Rev. Lett. 52 (1984) 489.

\bibitem{turok89}
N. Turok, Phys. Rev. Lett. 63 (1989) 2625.
\bibitem{ellis88}
J. Ellis and H. Kowalski, Phys. Lett. B214 (1988) 161.

\bibitem{degrand84}
T. A. DeGrand, Phys. Rev. D30 (1984) 2001.
\bibitem{ellis89}
J. Ellis, U. Heinz and H. Kowalski, Phys. Lett. B233 (1989) 223. 

\bibitem{glashow75}
A. De Rujula, H. Georgi and S. L. Glashow, Phys. Rev. D12 (1975) 147. 

\bibitem{beg64}
M. A. B. Beg, B. W. Lee and A. Pais, Phys. Rev. Lett. 13 (1964) 514.  

\end{thebibliography}
\end{document}